%% file: paper.tex
\documentclass[reprint,showkeys]{revtex4-2}
\usepackage{hyperref}
\usepackage{graphicx,bm}
\usepackage{amsmath,amssymb}
\begin{document}

\title{Fock Space and Field Theoretic Description of Nonequilibrium
  Work Relations}

\author{Andrew J. Baish}
\affiliation{Kyocera SLD Laser, 111 Castilian Dr, Goleta, CA 93117}

\author{Benjamin P. Vollmayr-Lee}

\affiliation{Department of Physics and Astronomy, Bucknell University,
  Lewisburg, PA 17837}

\date{\today}
\begin{abstract}
  We consider classical, interacting particles coupled to a thermal
  reservoir and subject to a local, time-varying potential while
  undergoing hops on a lattice.  We impose detailed balance on the
  hopping rates and map the dynamics to the Fock space Doi
  representation, from which we derive the Jarzynski and Crooks
  relations. Here the local potential serves to drive the system far
  from equilibrium and to provide the work.  Next, we utilize the
  coherent state representation to map the system to a Doi-Peliti
  field theory and take the continuum limit.  We demonstrate that time
  reversal in this field theory takes the form of a gauge-like
  transformation which leaves the action invariant up to a generated
  work term.  The time-reversal symmetry leads to a fundamental
  identity, from which we are able to derive the Jarzynski and Crooks
  relations, as well as a far-from-equilibrium generalization of the
  fluctuation-dissipation relation.
\end{abstract}


\maketitle


\section{Introduction}

In the past decades remarkable advances have been made in the field of
far from equilibrium statistical physics.  The celebrated Jarzynski
equation,
\begin{equation}
  \langle e^{-W/k_BT}\rangle = e^{-\Delta F/k_BT},
\end{equation}
relates via equality the difference in the equilibrium free energy
$\Delta F$ to a nonequilibrium average of the work $W$ for processes
that start in equilibrium at temperature $T$ but are driven
arbitrarily far from equilibrium.  This holds for Hamiltonian systems
\cite{Jarzynski1997} as well as stochastic systems described by a
master equation \cite{Jarzynski1997b}.  This universal constraint on
the possible distributions of nonequilibrium work was further
developed by Crooks, who showed that the probability $P_F(W)$ of doing
work $W$ while following the ``forward'' protocol (a specified
variation of control parameters) can be related to the probability
$P_R(-W)$ of doing negative the work during a reversed protocol via
\begin{equation}
  P_F(W) = e^{(W-\Delta F)/k_BT}P_R(-W),
\end{equation}
thus elucidating the central role of time reversal
\cite{Crooks1998,Crooks1999}.  The Jarzynski and Crooks relations are
part of an ensemble of fluctuation theorems that describe work
fluctuations and entropy production, with by now a large literature
that has been summarized in reviews
\cite{Jarzynski2008,Jarzynski2011,Seifert2012} and more recently in
textbooks \cite{Peliti2021,Shiraishi2023}.

In these studies one typically examines the impact of time reversal on
the probability distribution of all possible trajectories, although
this probability distribution is generally not available in explicit
form.  In the present work, we propose to recast the nonequilibrium
dynamics into a field theory.  While this necessarily represents a
loss of trajectory-level information, similar to a master equation
treatment, the field theory nevertheless offers advantages.  The
equivalent of the probability distribution for field, namely the
exponential of the action, can be obtained explicitly.  The explicit
expression for the action enables us to make progress on three fronts:
one, as with all field theories, it provides a useful framework for
exploring symmetries and near-symmetries, in this case, the behavior
under time reversal; two, it enables a variety of possible
approximation schemes by perturbation theory; and three, a hierarchy
of nonequilibrium identities can be derived by functional
differentiation, such as a nonequilibrium generalization of the
fluctuation-dissipation relation.

Mallick, Moshe, and Orland examined nonequilibrium work relations via
field theory for the critical dynamics of a nonconserved scalar order
parameter \cite{Mallick2011}, i.e. Model A in the Heisenberg-Haltering
scheme \cite{Hohenberg1977}.  They were able to derive the Jarzynski
and Crooks relations, as well as a nonequilibrium generalization of
the fluctuation-dissipation relation, and to elucidate properties of
time-reversal symmetry breaking with the field theory.  This work was
extended by T\"auber to the critical dynamics of a conserved scalar
order parameter, i.e., Model B \cite{Taeuber2014}.

In this work we present a field theoretic description of the Jarzynski
and Crooks relations for a system of interacting classical particles
coupled to the thermal reservoir and driven by a tun able local
potential.  We employ Doi-Peliti field theory, i.e., a mapping of the
master equation for the classical particle model to a Fock space
representation (the Doi representation \cite{Doi1976}) and then a
subsequent mapping to a field theory via coherent states
\cite{Peliti1985,Schulman1996,Taeuber2005}.  One of the advantages of
this approach is that we are not restricted to near-criticality
dynamics.

Our primary results are the following. After defining the model in
Sec.~\ref{sec:model}, we recast the dynamics in the Doi
representation in Sec.~\ref{sec:doi-representation}, deriving the
general form of a time-evolution operator that is consistent with
detailed balance.  With this formalism in
Sec.~\ref{sec:work-relations-in-doi-representation} we obtain a direct
and simple Fock space derivation of the Jarzynski and Crooks
relations.  Next, in Sec.~\ref{sec:doi-peliti-field-theory} we obtain
the Doi-Peliti field theory for interacting particles coupled to a
thermal bath.  This field theory contains boundary ``initial'' and
``final'' terms in the action as well as a bulk action governing the
time evolution from $t=0$ to $t=t_f$.  We demonstrate that the bulk
action can be transformed by a Cole-Hopf transformation to the
Dean-Kawasaki field theory
\cite{Andreanov2006b,Andreanov2006,Lefevre2007,Kim2014} obtained by
mapping Dean's Langevin equation for interacting particles
\cite{Dean1996} to a field theory via the
Martin-Siggia-Rose-Janssen-de Dominicis procedure
\cite{Martin1973,Janssen1976,deDominicis1976}.  This connection was
reported previously by Andreanov, et al. \cite{Andreanov2006b}, but
our treatment differs importantly in the boundary terms, which play a
role in time reversal.  We also demonstrate for noninteracting
particles that our field theory reduces to the Fokker-Planck equation
for a Brownian particle in a potential.

In Sec.~\ref{sec:time-reversal} we examine time reversal within the
field theory, showing that time-reversal takes the form of a
gauge-like transformation, and that the action is invariant under time
reversal apart from the introduction of the Jarzynski work term.  We
note that for interacting particles, it is necessary to introduce a
Hubbard-Stratonovich transformation and add a complex field to fully
reveal the underlying time reversal symmetry.  We obtain a fundamental
nonequilibrium identity, Eq.~(\ref{eq:fundamental-relation}), from
which the Jarzynski and Crook's identities follow. Further,
in Sec.~\ref{sec:nonequilibrium-identities} we employ functional
differentiation of Eq.~(\ref{eq:fundamental-relation}) to derive
a far-from-equilibrium generalization of the fluctuation-dissipation
relation.


\section{Model}

\label{sec:model}

\subsection{Energy and Thermal Equilibrium}

We begin with a model of $N$ classical particles on a $d$-dimensional
lattice of size $L^d$; the continuum limit will be taken later.  The
particles interact with the pair potential $V_{ij}$, which can be any
function of the separation ${\bf r}_j-{\bf r}_i$ between sites $i$ and
$j$, such as a Lennard-Jones potential, with the only restriction that
it is symmetric under interchange of particles, $V_{ij}=V_{ji}$.
Additionally, we consider a time-dependent site potential $U_i(t)$,
which allows for work to be done on the system.  The occupancy of site
$i$ is denoted $n_i$, and the energy of the occupation configuration
${\bf n}=(n_1,n_2,\dots)$ is given by
\begin{equation}
  E_{\bf n}(t) = \frac{1}{2}\sum_{i,j} (n_i-\delta_{ij}) V_{ij} n_j
  + \sum_i U_i(t) n_i,
  \label{eq:energy}
\end{equation}
where the Kronecker delta function ensures proper counting of the
same-site pairs. 

These particles are coupled to a thermal reservoir at temperature $T$.
Because of the time-dependent potential $U_i(t)$, the equilibrium
distribution also depends on time $t$ via
\begin{equation}
  P^{\rm eq}_{\bf n}(t) = \frac{N!}{{\bf n}!} \,\frac{1}{Z(t)}
  e^{-\beta E_{\bf n}(t)}.
  \label{eq:p-eq}
\end{equation}
where $\beta^{-1}=k_BT$, $Z(t)$ is the partition function associated
with $E_{\bf n}(t)$, and ${\bf n}!\equiv\prod_i n_i!$.  The prefactor
$N!/{\bf n}!$ is the multinomial coefficient that accounts for the
multiplicity of microstates for configuration ${\bf n}$ under
permutation of the particles.

\subsection{Dynamics}

The particles undergo stochastic hops, and the probability $P({\bf
  n},t)$ of configuration ${\bf n}$ at time $t$ obeys the master
equation
\begin{equation}
  \frac{d}{dt} P_{\bf n}(t) = \sum_{{\bf m}\neq {\bf n}}
  \Bigl[ w_{{\bf n},{\bf m}}(t)P_{\bf m}(t) -w_{{\bf m},{\bf n}}(t)P_{\bf n}(t)
    \Bigr].
  \label{eq:master-eq}
\end{equation}
The particle hopping rates $w_{{\bf n},{\bf m}}(t)$ for a transition
${\bf m}\to {\bf n}$ are chosen to obey detailed balance for the
instantaneous energy $E_{\bf n}(t)$,
\begin{equation}
  \frac{w_{{\bf n},{\bf m}}(t)}{w_{{\bf m},{\bf n}}(t)} = 
  \frac{P^{\rm eq}_{\bf n}(t)}{P^{\rm eq}_{\bf m}(t)} = \frac{{\bf m}!}{{\bf n}!}
  e^{-[\beta E_{\bf n}(t) - \beta E_{\bf m}(t)]}
  \label{eq:detailed-balance}
\end{equation}
and are thus time-dependent.  

Work is performed on the system by varying the site potentials
$U_i(t)$.  For configuration ${\bf n}$ at time $t$, the instantaneous
rate of work performed on the system is given by
\begin{equation}
  \dot W = \sum_i n_i \dot U_i(t).
  \label{eq:rate-of-work}
\end{equation}
This term appears in the nonequilibrium generalization of the first
law:
\begin{equation}
\frac{d}{dt}\langle E\rangle = \sum_{\bf n} \biggl[\frac{dP_{\bf n}(t)}{dt} 
E_{\bf n}(t) + P_{\bf n}(t)\sum_i n_i \dot U_i(t)\biggr].
\label{eq:nonequilibrium-first-law}
\end{equation}
The first term gives the rate of change in $\langle E\rangle$ due to
the particle hopping dynamics, reflecting the heat flow between the
system and reservoir, while the second term is simply the average rate
of work performed, $\langle \dot W\rangle$.

For sufficiently slow driving, the system remains in thermal
equilibrium and the probabilities are given by Eq.~(\ref{eq:p-eq}). In
contrast, for fast driving, the distribution $P_{\bf n}(t)$ obtained
from Eq.~(\ref{eq:master-eq}) may be driven far from equilibrium.


\section{The Doi Representation}

\label{sec:doi-representation}

These particle dynamics can be mapped to a Fock space representation,
i.e., the Doi representation \cite{Doi1976}, using a standard
procedure which we summarize briefly below.  Even though these are
classical particles, the Fock space representation reflects the
permutation symmetry of the particles, and thus provides in this sense
the simplest expression of the particle dynamics.  We shall see that,
indeed, in the Doi representation the Jarzynski relation emerges
elegantly.

\subsection{States and the Liouvillian}

Bosonic creation and annihilation operators $\hat a_i^\dagger$ and
$\hat a_i$ are introduced at each lattice site and, together with the
vacuum state $|0\rangle$, used to create a state $|{\bf n}\rangle =
\prod_i (\hat a_i^\dagger)^{n_i}|0\rangle$ corresponding to the
configuration ${\bf n}$.  Note that this state is normalized as as
$\langle{\bf n}|{\bf n}\rangle = {\bf n}!$.  The full probability
distribution of the system at time $t$ can then be represented by the
state
\begin{equation}
  |\psi(t)\rangle = \sum_{\bf n} P_{\bf n}(t) |{\bf n}\rangle,
  \label{eq:doi-state-def}
\end{equation}
which enables writing the master equation as
\begin{equation}
  \frac{d}{dt}|\psi(t)\rangle = -\hat L(t) |\psi(t)\rangle.
  \label{eq:doi-master-eq}
\end{equation}
with Liouvillian operator $\hat L$.  The utility of the Doi
representation is that whenever the particle dynamics respects the
permutation symmetry of the particles, the operator $\hat L$ can be
expressed solely in terms of the $\hat a_i$ and $\hat a_i^\dagger$
operators, with no dependence on the configuration ${\bf n}$.  That
is, the Liouvillian is determined solely by the processes involved and
not dependent on the state of the system.  For example, particles
undergoing unbiased hops between neighboring lattice sites leads to
\begin{equation}
  \hat L_{\rm diff} = \Gamma \sum_{\langle ij\rangle} 
   (\hat a_i^\dagger-\hat a_j^\dagger)(\hat a_i-\hat a_j),
   \label{eq:liouvillian-diffusion}
\end{equation}
where the sum runs over $i,j$ that are nearest neighbor sites.

\subsection{The Projection State}

In the Doi representation, averages and distributions are extracted
from the Fock states by means of a projection state, defined as
$\langle {\cal P}| = \langle 0 | e^{\sum_i \hat a_i}$, with the
property $\langle{\cal P}|\hat a_i^\dagger = \langle{\cal P}|$ for all
$i$.  Averages are computed via
\begin{equation}
  \langle A\rangle \equiv \sum_{\bf n} A_{\bf n} P_{\bf n}(t)
  = \langle {\cal P}|\hat A|\psi(t)\rangle,
  \label{eq:doi-averages}
\end{equation}
where $\hat A\equiv\sum_{\bf n}(A_{\bf n}/{\bf n}!) |{\bf
  n}\rangle\langle{\bf n}|$ is an operator diagonalized by the
occupation states $|{\bf n}\rangle$ with eigenvalues $A_{\bf n}$.
Normalization ensures $\langle{\cal P}|\psi(t)\rangle = \sum_{\bf
  n}P_{\bf n}(t) = 1$, and probability conservation requires
\begin{equation}
  0 = \frac{d}{dt}\langle 1\rangle 
  = -\langle {\cal P}|\hat L(t)|\psi(t)\rangle
\end{equation}
for all $|\psi(t)\rangle$, thus
\begin{equation}
  \langle {\cal P}|\hat L(t) = 0.
  \label{eq:probability-conservation}
\end{equation}
This is obeyed, for example, by the diffusion Liouvillian in
Eq.~(\ref{eq:liouvillian-diffusion}).

\subsection{The Hamiltonian and Detailed Balance}

Now we consider the class of Liouvillians that correspond to
transition rates that satisfy detailed balance, and for clarity we
suppress the explicit time dependence in $\hat L$ and the rates.
Starting from a state ${\bf m}$, the probability of being in a state
${\bf n}\neq {\bf m}$ a short time $\delta t$ later is, according to
the master equation, $P({\bf n},t) = w_{{\bf n},{\bf m}}\delta t$,
while the probability of remaining in ${\bf m}$ is $P_{\bf m}(\delta
t) = 1 - \delta t\sum_{{\bf n}'\neq {\bf m}} w_{{\bf n}',{\bf m}}$.
In the Doi representation, the state at time $\delta t$ is
$|\psi(\delta t)\rangle = (1 - \hat L\delta t) |{\bf m}\rangle$.
Using $\langle{\bf n}|\psi(t)\rangle = {\bf n}! P_{\bf n}(t)$ we
obtain for ${\bf n}\neq {\bf m}$
\begin{equation}
  \langle{\bf n}|\hat L|{\bf m}\rangle 
  = -{\bf n}! \, w_{{\bf n},{\bf m}} 
  \label{eq:Lmn-def}
\end{equation}
while for ${\bf n} = {\bf m}$
\begin{equation}
  \langle{\bf n}|\hat L|{\bf n}\rangle = -{\bf n}! \sum_{{\bf n}'\neq {\bf n}}
    w_{{\bf n}',{\bf n}}.
    \label{eq:Lnn-def}
\end{equation}
For rates that obey the detailed balance condition of
Eq.~(\ref{eq:detailed-balance}), we obtain
\begin{equation}
  \langle{\bf n}|\hat L|{\bf m}\rangle =
  \langle{\bf m}|\hat L|{\bf n}\rangle e^{-\beta E_{\bf n}(t) + \beta E_{\bf m}(t)}.
  \label{eq:liouvillian-me-detailed-balance}
\end{equation}
with $E_{\bf n}$ given by Eq.~(\ref{eq:energy}).  Now we introduce the
hermitian Hamiltonian operator
\begin{equation}
  \hat H(t) = \frac{1}{2}\sum_{i,j} V_{ij} \hat a_i^\dagger \hat a_j^\dagger
  \hat a_j \hat a_i + \sum_i U_i(t) \hat a_i^\dagger \hat a_i
  \label{eq:hamiltonian}
\end{equation}
which is diagonalized by the states $|{\bf n}\rangle$ with eigenvalues
$E_{\bf n}(t)$. With this energy operator
Eq.~(\ref{eq:liouvillian-me-detailed-balance}) becomes $\langle{\bf
  n}|\hat L|{\bf m}\rangle = \langle{\bf m}|e^{\beta\hat H}\hat
Le^{-\beta\hat H}|{\bf n}\rangle$, which implies
\begin{equation}
  \hat L^\dagger = e^{\beta\hat H}\hat L e^{-\beta\hat H}.
  \label{eq:detailed-balance-for-L}
\end{equation}
Together Eqs.~(\ref{eq:probability-conservation}) and
(\ref{eq:detailed-balance-for-L}) imply that necessary and sufficient
conditions for detailed balance are that $\hat L$ has the form
\begin{equation}
  \hat L = \hat Q e^{\beta\hat H}
  \label{eq:liouvillian-detailed-balance}
\end{equation}
where $\hat Q$ is a hermitian operator that annihilates the projection
state: $\hat Q|{\cal P}\rangle=0$.

\subsection{Constructing the Liouvillian}

Finally, we discuss specific choices for the Liouvillian. Consider a
particle hop from site $i$ to neighboring site $j$, which takes the
system from configuration ${\bf m}$ to ${\bf n}$, where $n_k = m_k
-\delta_{ik} +\delta_{jk}$.  The detailed balance condition
(\ref{eq:detailed-balance}) for this hop is
\begin{equation}
  \frac{w_{{\bf n},{\bf m}}}{w_{{\bf m},{\bf n}}} = 
\frac{m_i}{n_j} e^{-\beta [E_{\bf n}-E_{\bf m}]}.
\label{eq:detailed-balance-condition}
\end{equation}
We restrict consideration to rates that depend on the energies only
via the difference $\Delta E\equiv E_{\bf n}-E_{\bf m}$.  These have
the general form
\begin{equation}
w_{{\bf n},{\bf m}} = \Gamma m_i f_e(\beta\Delta E) e^{-\beta \Delta E/2},
\label{eq:detailed-balance-rates}
\end{equation}
where $f_e(x)$ is an even function.

To construct the Liouvillian corresponding to
(\ref{eq:detailed-balance-rates}) it is useful to introduce the
hermitian operator
\begin{equation}
\hat\epsilon_k = U_k + \sum_{\ell} V_{k\ell} \hat a_\ell^\dagger \hat a_\ell
\end{equation}
with eigenvectors $|{\bf m}\rangle$ and eigenvalues $\epsilon_k({\bf
  m}) = U_k + \sum_\ell V_{k\ell} m_\ell$ equal to the energy required
to introduce a particle at site $k$ to configuration ${\bf m}$.  It is
straightforward to show that for the $i\to j$ jump
\begin{equation}
(\hat\epsilon_j-\hat\epsilon_i)\,\hat a_i|{\bf m}\rangle
  = \Delta E\, \hat a_i|{\bf m}\rangle,
\end{equation}
which allows us to write (\ref{eq:detailed-balance-rates}) as
\begin{equation}
 w_{{\bf n},{\bf m}} = \frac{1}{{\bf n}!}
\langle{\bf n}| \Gamma \hat a_j^\dagger f_e(\hat\epsilon_i-\hat\epsilon_j) 
  e^{\beta(\hat\epsilon_i-\hat\epsilon_j)/2} 
\hat a_i |{\bf m}\rangle.
\label{eq:L-gain-term}
\end{equation}
By comparison with (\ref{eq:Lmn-def}), we can identify the operator
above with the gain term in $\hat L$.

For the loss term, we consider the same $i\to j$ hop but starting in
${\bf n}$ and going to ${\bf n}'$, where
$n_k'=n_k-\delta_{ik}+\delta_{jk}$, and $\Delta E=E_{{\bf n}'}-E_{\bf
  n}$.  The appropriate rate is obtained from (\ref{eq:Lnn-def}) and
(\ref{eq:detailed-balance-rates}) to be
\begin{equation}
  w_{{\bf n}\to{\bf n}'} = \frac{1}{{\bf n}!} 
  \langle{\bf n}| \Gamma \hat a_i^\dagger f_e(\hat\epsilon_i-\hat\epsilon_j) 
   e^{\beta(\hat\epsilon_i-\hat\epsilon_j)/2} 
  \hat a_i |{\bf n}\rangle.
  \label{eq:L-loss-term}
\end{equation}
Combining (\ref{eq:L-gain-term}) and (\ref{eq:L-loss-term}) with the
analogous terms for a $j\to i$ hop provides the Liouvillian
corresponding to (\ref{eq:detailed-balance-rates}),
\begin{equation}
\hat L = \Gamma \sum_{\langle ij\rangle} (\hat a_i^\dagger-\hat a_j^\dagger)
f_e(\hat\epsilon_i-\hat\epsilon_j)
\Bigl(e^{\beta(\hat\epsilon_i-\hat\epsilon_j)/2}\hat a_i-
e^{\beta(\hat\epsilon_j-\hat\epsilon_i)/2}\hat a_j\Bigr).
\label{eq:liouvillian}
\end{equation}
As advertised, the Liouvillian is purely an expression of the process
and has no dependence on the state of the system, i.e., the occupation
numbers.  The identity $e^{\beta\hat\epsilon_k} \hat a_k =
e^{-\beta\hat H} \hat a_k e^{\beta\hat H}$, which follows from $[\hat
  H,\hat a_k] = -\hat\epsilon_k \hat a_k$ and the Hadamard lemma, can
be used to show this Liouvillian has the necessary form
(\ref{eq:liouvillian-detailed-balance}).

Later, when taking the continuum limit, the energy difference due to a
hop will be of the order of the lattice spacing $\Delta x$, which will
allow us to linearize (\ref{eq:liouvillian}) in
$\hat\epsilon_i-\hat\epsilon_j$:
\begin{equation}
  \hat L \simeq \Gamma\sum_{\langle ij\rangle} (\hat a_i^\dagger-\hat
  a_j^\dagger) \biggl(\hat a_i-\hat a_j +
  \beta(\hat\epsilon_i-\hat\epsilon_j)\frac{\hat a_i+\hat a_j}{2}
  \biggr),
\label{eq:linearized-liouvillian}
\end{equation}
where we have set $f_e(0)=1$ without loss of generality.  Note that
this linearized Liouvillian is normal ordered.

\subsection{Equilibrium State}

The Jarzynski and Crooks relations require starting in thermal
equilibrium.  In the Doi representation the equilibrium state can be
written as
\begin{equation}
  |\psi_\text{eq}\rangle = \sum_{\bf n} P_{\bf n}^\text{eq} |{\bf n}\rangle
  = \frac{1}{Z} e^{-\beta\hat H} |\bar n_0\rangle,
  \label{eq:equilibrium-state}
\end{equation}
where $\bar n_0 = N/L^d$ is the average number of particles per site,
and $|\bar n_0\rangle \equiv e^{\sum_i \bar n_0 \hat
  a_i^\dagger}|0\rangle$.  This is obtained from (\ref{eq:p-eq}) by
expressing the multiplicity of the occupation state ${\bf n}$ in terms
of a product of Poisson distributions: $N!/{\bf n}! \propto \prod_i
\bar n_0^{n_i}/n_i!$, with the implied constraint $\sum_i n_i =N$ and
the proportionality constant absorbed into $Z$.

The equilibrium state is necessarily a stationary state of the
detailed balance Liouvillian,
Eq.~(\ref{eq:liouvillian-detailed-balance}).  In the Doi
representation this can be seen from
\begin{equation}
  \hat L |\psi_\text{eq}\rangle = Z^{-1} \hat Q |\bar n_0\rangle = 0,
  \end{equation}
where the last equation follows from taking $\hat a \to \bar n_0 \hat
a$ and $\hat a^\dagger \to \bar n_0^{-1} \hat a^\dagger$, which leaves
$\hat Q$ invariant and takes $|\bar n_0\rangle \to |{\cal P}\rangle$.

Finally, we note that all of $\hat L$, $\hat H$, $E$, $Z$, and
$|\psi_\text{eq}\rangle$ are defined in terms of the instantaneous
$U_i(t)$, and thus can be time-dependent.


\section{Work Relations in the Doi Representation}

\label{sec:work-relations-in-doi-representation}

We are now equipped to derive the Jarzynski relation for general
detailed balance dynamics within the Doi representation. Perhaps
unsurprisingly, our methods bear a strong similarity to the operator
framework Kurchan used for Langevin dynamics \cite{Kurchan2007}. As
usual, the system must begin in thermal equilibrium at time $t=0$; we
denote this state $|\psi_\text{eq}(0)\rangle$ to emphasize that the
time-dependent Hamiltonian is to be evaluated at $t=0$.

The solution to the master equation (\ref{eq:doi-master-eq}) can be
written as
\begin{equation}
  |\psi(t)\rangle = \lim_{\Delta t\to 0} \prod_{n=0}^{t/\Delta t-1} 
  (1 -\hat L_{t_n}\Delta t) |\psi_\text{eq}(0)\rangle.
  \label{eq:time-slice-solution}
\end{equation}
where the product of the time-dependent $\hat L_t$ operators is time
ordered with earlier times on the right, and $t_n = n\Delta t$.  Work,
as defined in (\ref{eq:rate-of-work}), appears in the Doi representation
as
\begin{equation}
  W = \int_0^{t_f} dt\, \sum_i \dot U_i(t) \hat a_i^\dagger \hat a_i.
  \label{eq:doi-work}
\end{equation}
The
average of the work from time $t=0$ to $t_f$ is given by
\begin{equation}
  \langle e^{-\beta W}\rangle 
  = \langle {\cal P}| \prod_{n=0}^{n_f} \left[
e^{-\beta \dot W_{t_n}\Delta t}(1-\hat L_{t_n}\Delta t)\right]
  |\psi_\text{eq}(0)\rangle
  \label{eq:jarzynski-average-v1}
\end{equation}
where $n_f= t_f/\Delta t - 1$, the product is again time ordered, and
the limit $\Delta t\to 0$ is implied.
Eq.~(\ref{eq:jarzynski-average-v1}) is equivalent to applying a weight
$e^{-\beta w(t)}$ to each trajectory, where $w(t)$ is the work done up
to time $t$ along that trajectory.  This has been demonstrated to
provide the desired average \cite{Jarzynski1997b,Hummer2001}.

Substituting $e^{-\beta \dot W_t\Delta t}=e^{-(\beta \hat H_{t+\Delta
    t}-\beta\hat H_t)}$ and regrouping terms gives
\begin{align}
  \langle e^{-\beta W}\rangle 
  = \langle {\cal P}|& e^{-\beta\hat H_{t_f}}\prod_{n=0}^{n_f} \Bigl[e^{\beta\hat H_{t_n}}
    (1-\hat L_{t_n}\Delta t) e^{-\beta\hat H_{t_n}}\Bigr] 
  \nonumber\\
  &\times  e^{\beta\hat H_{0}} |\psi_\text{eq}(0)\rangle.
  \label{eq:jarzynski-average-v2}
\end{align}
Utilizing the detailed balance condition
(\ref{eq:detailed-balance-for-L}), the square brackets become
\begin{equation}
  e^{\beta\hat H_t}(1-\hat L_{t}\Delta t) e^{-\beta\hat H_{t}}  = 
  1 - \hat L_t^\dagger \Delta t,
\end{equation}
which can be interpreted as simply the time evolution operator acting
to the left.  This is the essence of how the fluctuation relations
appear in the Doi representation: a forward-time average including the
work term transforms into the reverse-time average absent the work.

It remains to analyze the initial and final states. We rescale the
operators $\hat a \to \bar n_0 \hat a$ and $\hat a^\dagger \to \bar
n_0^{-1} \hat a^\dagger$, which leaves $\hat L$ and $\hat H$ (and
therefore the work term) unchanged, but modifies the initial and final
terms via
\begin{equation}
  e^{\beta \hat H_0}|\psi_\text{eq}(0)\rangle = Z(0)^{-1} |\bar n_0\rangle
  \to Z(0)^{-1} |{\cal P}\rangle
\end{equation}
and
\begin{equation}
  \langle{\cal P}| e^{-\beta H_{t_f}}\to \langle\bar n_0|e^{-\beta H_{t_f}}
  = Z(t_f)^{-1}\langle\psi_\text{eq}(t_f)|
\end{equation}
Putting this together, we have
\begin{equation}
  \langle e^{-\beta W}\rangle 
  = \frac{Z(t_f)}{Z(0)}\langle\psi_\text{eq}(t_f)|\prod_{n=0}^{t_f/\Delta t-1}
  (1-\hat L^\dagger_{t_n}\Delta t) |{\cal P}\rangle
  \label{eq:reversed-trajectory}
\end{equation}
Conjugating the real expectation value reverses the time ordering,
giving
\begin{equation}
  \langle e^{-\beta W}\rangle 
  = \frac{Z(t_f)}{Z(0)}\langle {\cal P}|
  (1-\hat L_{0}\Delta t)\dots
  (1-\hat L_{t_f-\Delta t}\Delta t)|\psi_\text{eq}(t_f)\rangle_R
\end{equation}
which can be interpreted as time ordered in the variable $t'=t_f-t$.
The expectation value corresponds to $\langle {\cal P} |
\psi(t'=t_f)\rangle_R = 1$ and we obtain the Jarzynski relation
\begin{equation}
  \langle e^{-\beta W}\rangle 
  = \frac{Z(t_f)}{Z(0)} = e^{-\beta[\Delta F(t_f)-F(0)]}.
\end{equation}

To derive the Crooks relation, consider the characteristic function
$\phi_F(q)$ of the forward work distribution function, which is simply
the Fourier transform:
\begin{equation}
  \phi_F(q) = \int_{-\infty}^\infty dW\, e^{iq W} P_F(W) = \langle
  e^{iqW}\rangle_F.
\end{equation}
This takes the form of Eqs.~(\ref{eq:jarzynski-average-v1}) but
with $\beta \to iq$.  We now have a repeating pattern of
\begin{equation}
  e^{-iq\hat H_t}(1-\hat L_{t}\Delta t) e^{iq\hat H_{t}}  =
  e^{-(iq+\beta)\hat H}(1 - \hat L_t^\dagger \Delta t) e^{(iq+\beta)\hat H}
\end{equation}
which means after time reversal
\begin{equation}
  \langle e^{iqW}\rangle_F = \frac{Z(t_f)}{Z(0)}
  \langle e^{-iqW}e^{-\beta W}\rangle_R
  \label{eq:crooks}
\end{equation}
or $\phi_F(q) = \phi_R(iW-q) e^{-\beta\Delta F}$.  Inverse
transforming then gives
\begin{equation}
  P_F(W) = \int \frac{dq}{2\pi} e^{-iqW}\phi_R(iW-q) = e^{\beta W} P_R(-W),
\end{equation}
which is the Crooks relation.


\section{Doi-Peliti Field Theory}

\label{sec:doi-peliti-field-theory}

For taking the limit from a lattice to a spatial continuum, it is
desirable to convert the Doi representation of the dynamics to a field
theory.  In this section we use the coherent state representation
\cite{Schulman1996} to derive the Doi-Peliti field theory for the
particle model with detailed balance dynamics. The general technique
has been presented elsewhere \cite{Peliti1985,Lee1994,Taeuber2005}, so
we provide only a brief sketch here.

\subsection{Coherent State Representation}

Coherent states are introduced at each lattice site,
\begin{equation}
  |{\bm\phi}\rangle = e^{-\frac{1}{2}\sum_i |\phi_i|^2
    + \sum_i \phi_i \hat a_i^\dagger} |0\rangle
\end{equation}
with ${\bm\phi} = (\phi_1,\phi_2,\dots)$ and complex $\phi_i$.  These
are eigenstates of the annihilation operator, $a_i|{\bm\phi}\rangle
 = \phi_i |{\bm\phi}\rangle$. The
identity operator can be expressed with the overcompleteness relation
$\bm{1} = \int \prod_i \frac{d^2\phi_i}{\pi}|{\bm\phi}\rangle
\langle{\bm\phi}|$.  Time evolution is broken into discrete steps of
size $\Delta t$, as shown in (\ref{eq:time-slice-solution}), and the
identity operator is inserted at each time slice with a distinct set
of coherent states, leading to evaluation of terms of the form
\begin{equation}
  \langle {\bm\phi}_{t+\Delta t}| 1 - \hat L_t\Delta t |
          {\bm\phi}_t \rangle =\langle\bm\phi_{t+\Delta t} | \bm\phi_t\rangle 
  [1-\Delta t {\cal L}(\bm{\phi^*}_{t+\Delta t},\bm\phi_t)].
  \label{eq:matrix-element}
\end{equation}
Using the linearized, normal ordered Liouvillian
(\ref{eq:linearized-liouvillian}) we obtain
\begin{widetext}
\begin{equation}
  {\cal L}(\bm{\phi^*}_{t+\Delta t},\bm\phi_t)=\Gamma
  \sum_{\langle ij\rangle} (\phi^*_{i,t+\Delta t}-\phi^*_{j,t+\Delta t})
  \left(\phi_{i,t}-\phi_{j,t} 
  + \frac{\phi_{i,t}+\phi_{j,t}}{2}  
  \left\{\beta\epsilon_i({\bm\phi}_{t+\Delta t},{\bm\phi}_t) - 
  \beta\epsilon_j({\bm\phi}_{t+\Delta t},{\bm\phi}_{t})\right\}
  \right)
\end{equation}
\end{widetext}
with the effective potential
\begin{equation}
  \epsilon_i({\bm\phi}_{t+\Delta t}) = U_i +\sum_k V_{ik}\phi^*_{k,t+\Delta t}
  \phi_{k,t}.
\end{equation}
The coherent state overlap in (\ref{eq:matrix-element})  can be written as
\begin{equation}
  \langle\bm\phi_{t+\Delta t}|\bm\phi_t\rangle
  = e^{\frac{1}{2}|\bm\phi_{t+\Delta t}|^2  
  -\frac{1}{2}|\bm\phi_{t}|^2 
- \bar{\bm\phi}_{i,t+\Delta t}\cdot(\bm\phi_{t+\Delta t}-\bm\phi_{t})}
  \label{eq:overlap-relation}
\end{equation}
where $|{\bm\phi}|^2 = \sum_i|\phi_i|^2$ and
$\bar{\bm\phi}_1\cdot\bm\phi_2 =\sum_i \phi^*_{1,i}\phi_{2,i}$.  The
squared terms cancel between successive time slices.

We take the continuum limit via $\phi_{i,t}\to \phi({\bf x},t)\Delta
x^d$ and $\phi^*_{i,t}\to \bar\phi({\bf x},t)$, so $\phi({\bf x})$ has
dimensions of density while $\bar\phi({\bf x})$ is non-dimensional.
The result is an action $S$ containing a ``bulk'' contribution as well
as initial and final contributions at $t=0$ and $t_f$, which can be
used to compute averages via
\begin{equation}
  \langle A\rangle = \int {\cal D}(\bar\phi,\phi) \,
  A(\bar\phi,\phi) e^{-S[\bar\phi,\phi]}
  \label{eq:dp-average}
\end{equation}

\subsection{Doi-Peliti Action for Interacting Particles}

From Eqs.~(\ref{eq:matrix-element}) and (\ref{eq:overlap-relation}) we
obtain the bulk action
\begin{widetext}
\begin{equation}
  S_B = \int_{0}^{t_f}dt\int d{\bf x} 
  \biggl[\bar\phi(\partial_t - D\nabla^2)\phi 
    - \gamma\bar\phi\nabla\cdot\left(\phi\nabla U
    + \phi\nabla \int d{\bf y} V({\bf x}-{\bf y})\bar\phi({\bf y})
    \phi({\bf y})\right)\biggr]
  \label{eq:dp-bulk-action}
\end{equation}
\end{widetext}
with diffusion constant $D = \Gamma\Delta x^2$ and mobility
$\gamma=D/k_BT$.

Note that for noninteracting particles, with $V=0$, the action
(\ref{eq:dp-bulk-action}) is linear in $\bar\phi$, which creates a
delta function when integrated over $\bar\phi$.  The resulting field
$\phi$ then obeys
\begin{equation}
  \frac{\partial\phi}{\partial t} = D\nabla^2\phi + \gamma \nabla
  \cdot (\phi\nabla U)
\end{equation}
which is the Fokker-Planck equation for particles diffusing in a potential
$U({\bf x})$.

The final term, consisting of the projection state and part of the
coherent state overlap, contributes a $t_f$ ``boundary'' contribution
to the action of the form
\begin{equation}
  e^{-S_f} = e^{\frac{1}{2}|{\bm\phi}_{t_f}|^2}\langle{\cal P} | {\bm\phi}\rangle
  = e^{\sum_i\phi_{i,t_f}}
\end{equation}
from which we obtain the continuum limit
\begin{equation}
  S_f = - \int d{\bf x}\; \phi({\bf x},t_f).
  \label{eq:Sf}
\end{equation}
Generally this term is eliminated by performing a Doi shift.  Instead,
we will retain this term, since it plays an essential role in
understanding the time-reversal symmetry.

The initial term also provides a $t=0$ boundary term
\begin{equation}
  e^{-S_i} = Z(0)^{-1} e^{-\frac{1}{2}|\bm\phi_0|^2}\langle\bm\phi_0|
  e^{-\beta\hat H_0}|\bar n_0\rangle,
  \label{eq:Si-def}
\end{equation}
In the case of noninteracting particles, where $\hat H = \sum_j
U_j\hat a_j^\dagger \hat a_j$, this becomes
\begin{equation}
  e^{-S_i} = Z(0)^{-1} e^{-|\bm\phi_0|^2}\exp\left(\sum_j e^{-\beta U_{j,0}}\phi^*_{j,0}
  \bar n_0\right)
  \label{eq:Si-step-two}
\end{equation}
where we used the coherent state identity $\langle \phi_2 | e^{\lambda
  \hat a^\dagger \hat a} | \phi_1\rangle = \exp\{ (e^\lambda
-1)\phi_2^*\phi_1 \} \langle \phi_2 | \phi_1 \rangle$
\cite{Katriel2000}.  In the continuum limit, we take $\bar n_0\to
n_0\Delta x^d$ and obtain
\begin{equation}
  S_i = \ln Z(0) - \int d{\bf x} \left[ \bar\phi({\bf x},0)
    e^{-\beta U({\bf x},0)}n_0 - \bar\phi({\bf x},0)\phi({\bf x},0)\right]
  \label{eq:Si-noninteracting}
\end{equation}
To derive a simple expression for the initial
action with interacting particles, it will be necessary to introduce a
Hubbard-Stratonovich transformation, which we will describe below.

Finally, we note that in the field theory the work term (\ref{eq:doi-work})
becomes
\begin{equation}
  W[\bar\phi,\phi] = \int_0^{t_f} dt \int d{\bf x}\,  \dot U(t)
  \bar\phi({\bf x},t) \phi({\bf x},t).
  \label{eq:dp-work}
\end{equation}

\subsection{Cole-Hopf Transformation and Dean-Kawasaki Field Theory}

\label{subsec:DKfield_theory}

The bulk action (\ref{eq:dp-bulk-action}) can be transformed
via the Cole-Hopf transformation $\phi\to e^{-\bar\rho}\rho$,
$\bar\phi\to e^{\bar\rho}$ to
\begin{align}
  S_B = \int_{0}^{t_f}dt\int d{\bf x} 
  \biggl[\bar\rho(\partial_t - D\nabla^2)\rho 
    - \gamma\bar\rho\nabla\cdot(\rho\nabla \epsilon)
     - \gamma D\rho(\nabla\bar\rho)^2\biggr]
  \label{eq:dp-bulk-action2}
\end{align}
with the effective potential
\begin{equation}
  \epsilon({\bf x}) = U({\bf x}) + \int d{\bf y} V({\bf x}-{\bf y})
  \rho({\bf y}).
\end{equation}
This matches the Dean-Kawasaki field theory \cite{Kim2014} obtained
from the Langevin equation for interacting particles derived by Dean
\cite{Dean1996}, then mapped to a field theory using the MSRJD
technique \cite{Martin1973,Janssen1976,deDominicis1976}.  However,
there appear to be discrepancies with the initial and final terms.  In
particular, Dean-Kawasaki field theory has no analog of the projection
state and final action $S_f$.  Further the Cole-Hopf transformation
does not map our initial action to the generally assumed initial
conditions for Dean-Kawasaki field theory.

\section{Time Reversal in Doi-Peliti Field Theory}

\label{sec:time-reversal}

We are now equipped to examine the role of time reversal in the field
theory.  We begin with noninteracting particles, where the symmetry is
more directly manifest, and then introduce a Hubbard-Stratonovich
transformation to  reveal the symmetry for interacting particles.

\subsection{Noninteracting Particles}

When $V=0$, the bulk action (\ref{eq:dp-bulk-action}) can be written as
\begin{equation}
  S_B = \int_0^{t_f} dt\int d{\bf x} \biggl[ \bar\phi\partial_t \phi
    + D e^{-\beta U} \nabla\bar\phi \cdot \nabla (\phi e^{\beta U})\biggr]
  \label{eq:bulk-action-noninteracting}
\end{equation}
with $S_i$ and $S_f$ given by Eqs.~(\ref{eq:Si-noninteracting}) and
(\ref{eq:Sf}).  The time-reversal symmetry of the action is revealed
by the transformation
\begin{equation}
  \phi \to n_0 e^{-\beta U}\bar\psi \qquad
  \bar\phi \to n_0^{-1} e^{\beta U}\psi \qquad
  t \to t' = t_f-t
  \label{eq:transformation-noninteracting}
\end{equation}
This gauge-like transformation maintains the bilinear product
$\bar\phi\phi \to \bar\psi\psi$ and the action
(\ref{eq:bulk-action-noninteracting}) is invariant apart from the
time-derivative term.  Suppressing the spatial integral, we have
\begin{equation}
  \int_0^{t_f} dt\, \bar\phi\partial_t \phi  \to
  (\bar\psi\psi)\bigg\vert_{t=0}^{t=t_f} + \int_0^{t_f}dt'\biggl(
  \bar\psi\partial_{t'} \psi + \beta\frac{\partial U}{\partial t'}
  \bar\psi\psi\biggr)
  \label{eq:transformation-time-derivative}
\end{equation}
where we have used integration by parts.  Note that time reversal
generates a work term (\ref{eq:dp-work}).

The time reversal
transformation~(\ref{eq:transformation-noninteracting}) also maps the
initial action (\ref{eq:Si-noninteracting}) into the final action
(\ref{eq:Sf}) and vice-versa, with the help of the boundary terms in
(\ref{eq:transformation-time-derivative}).  The one exception is the
partition function factor, which remains $Z(0)^{-1}$ although the time-reversed
averaging would require that we start at $t'=0$ with $Z(t=t_f)^{-1}$.
Adjusting for the partition function and combining the bulk, initial,
and final actions, we obtain
\begin{equation}
  S[\bar\phi,\phi]  \to S_R[\bar\psi,\psi]
  + \beta W_R + \beta F(t=t_f) - \beta F(t=0),
  \label{eq:time-reversed-action}
\end{equation}
where $S_R$ and $W_R$ are the time-reversed action and work term, and
$\beta F = -\ln Z$.

As a consequence of this time reversal symmetry we have the identity
\begin{align}
  \langle A e^{-\beta W}\rangle &= \int {\cal D}(\bar\phi,\phi)
  A[\bar\phi,\phi] e^{-\beta
    W[\bar\phi,\phi]} e^{-S[\bar\phi,\phi]} \nonumber\\
  &\to
  e^{-\beta\Delta F} \int {\cal D}(\bar\psi,\psi) A_R[\bar\psi,\psi]
  e^{-S_R[\bar\psi,\psi]}\nonumber\\
  &= e^{-\beta\Delta F} \langle A_R \rangle_R
  \label{eq:fundamental-relation}
\end{align}
for any observable $A[\bar\phi,\phi]$, with time-reversed
$A_R[\bar\psi,\psi]$.
Here we have used that work (\ref{eq:dp-work}) is odd under time
reversal, $W[\bar\phi,\phi] \to -W_R[\bar\psi,\psi]$, to move the
exponential of negative work to the forward time average.  We will
show Eq.~(\ref{eq:fundamental-relation}) holds for interacting particles
as well, and provides the basis for deriving multiple nonequilibrium
identities.

In particular, setting $A=1$ produces the Jarzynski relation, while taking
$A = e^{(iq+\beta)W}$ reproduces Eq.~(\ref{eq:crooks}), which can be
inverse Fourier transformed to give the Crooks relation.

\subsection{Interacting Particles}

For interacting particles the bulk action (\ref{eq:dp-bulk-action})
can be written as
\begin{equation}
  S_B = \int_0^{t_f} dt\int d{\bf x} \biggl[ \bar\phi\partial_t \phi +
    D e^{-\beta \epsilon} \nabla\bar\phi \cdot \nabla (\phi e^{\beta
      \epsilon})\biggr]
  \label{eq:bulk-action-interacting}
\end{equation}
where $\epsilon({\bf x},t)$ is the effective potential, given by
\begin{equation}\epsilon({\bf x},t) = U({\bf x},t) + \int d{\bf y}\,
V({\bf x}-{\bf y}) \bar\phi({\bf y},t) \phi({\bf y},t).
\end{equation}
To affect time reversal with a local transformation like
(\ref{eq:transformation-noninteracting}), we need to employ a field
Hubbard-Stratonovich transformation:
\begin{align}
  \exp&\biggl(\beta\int d{\bf y}V({\bf x}-{\bf y})\bar\phi({\bf y})\phi({\bf y})
  \biggr)
  \nonumber\\
  &= A\int{\cal D}\eta  \exp\biggl\{{-\frac{1}{2\beta}}\int d{\bf r}_1
    d{\bf r}_2\,
    \eta^*({\bf r}_1) V^{-1}({\bf r}_{12})\eta({\bf r}_2)
    \nonumber\\
    &\quad + \int d{\bf r}_1\, \eta({\bf r}_1)\bar\phi({\bf r}_1)\phi({\bf r}_1)
    \biggr\} e^{-i\eta_2({\bf x})}
    \nonumber\\
    &\equiv \int {\cal D}\eta F[\eta,\bar\phi\phi] e^{-i\eta_2({\bf x})}
    \label{eq:hubbard-stratonovich}
\end{align}
where $\eta({\bf x},t) = \eta_1({\bf x},t)+ i\eta_2({\bf x}.t)$ is a
complex field, $A$ is a normalization constant, ${\bf r}_{12} = {\bf
  r}_1-{\bf r}_2$, and $V^{-1}$ is defined formally via $\int d{\bf y}
V({\bf x}-{\bf y}) V^{-1}({\bf y}-{\bf z}) = \delta^{(d)}({\bf x}-{\bf
  z})$. Eq.~(\ref{eq:hubbard-stratonovich}) applies for all times $t$,
with $\eta({\bf r},t)$ uncorrelated for different times.  With this
notation, the bulk action becomes
\begin{equation}
  S_B = \int{\cal D}\eta \,F
  \int dt d{\bf x} \biggl[ \bar\phi\partial_t \phi +
    D e^{-\beta \epsilon} \nabla\bar\phi \cdot \nabla (\phi e^{\beta
      U - i\eta_2})\biggr]
  \label{eq:bulk-action-hs}
\end{equation}
where we have suppressed the time integration limits.  Now the
gauge-like field
transformation
\begin{equation}
  \phi \to n_0 e^{-\beta U+i\eta_2}\bar\psi \qquad
  \bar\phi \to n_0^{-1} e^{\beta U-i\eta_2}\psi
  \label{eq:transformation-interacting}
\end{equation}
with time reversal $t \to t' = t_f-t$ once again leaves the action
invariant, i.e.,
\begin{equation}
  \nabla\bar\phi \cdot \nabla(\phi e^{\beta U - i\eta_2})
  \to \nabla(\psi e^{\beta U-i\eta_2})\cdot\nabla\bar\psi
\end{equation}
apart from the time derivative term, which again has the form given in
Eq.~(\ref{eq:transformation-time-derivative}), generating the work
term.  Importantly, the time derivative on the $\eta$ field averages
to zero because the field is uncorrelated over time.

The Hubbard-Stratonovich transformation also simplifies the initial
action.  The exponential of the Doi hamiltonian (\ref{eq:hamiltonian})
can be written as
\begin{equation}
  e^{-\beta\hat H} = A\int\prod_i d\eta_{i} e^{-\frac{1}{2\beta}
    \sum_{jk} \eta_{j} V^{-1}_{jk}\eta_{k} + \sum_j (-\beta U_j + i\eta_{j})
    \hat a_j^\dagger \hat a_j}.
\end{equation}
Using this in our initial action $S_i$, as defined in
Eq.~(\ref{eq:Si-def}), we obtain an identical expression to
(\ref{eq:Si-step-two}) with $e^{-\beta U_{j,0}}\to e^{-\beta
  U_{j,0}+i\eta_j}$ and the additional integration over the $\eta_j$
variables.  Taking the continuum limit, with $\eta_j \to \eta_2({\bf x},0)$,
results in an initial action of the form
\begin{align}
  S_i = \ln Z(0) - \int d\eta F \int d{\bf x}\Bigl[& \bar\phi({\bf x},0)
    e^{-\beta U({\bf x},0)+i\eta_2({\bf x},0)}n_0  \nonumber\\
    &- \bar\phi({\bf x},0)\phi({\bf x},0)
    \Bigr]
\end{align}
while the final action (\ref{eq:Sf}) remains unchanged.  Thus the
field transformation (\ref{eq:transformation-interacting}) swaps the
initial and final actions apart from the partition function, exactly
as before with the noninteracting theory.  Thus the fundamental identity
(\ref{eq:fundamental-relation}) continues to hold for interacting particles.

\section{Nonequilibrium Identities}

\label{sec:nonequilibrium-identities}

Starting from the fundamental relation (\ref{eq:fundamental-relation})
we can derive a number of nonequilibrium identities by appropriate
choice of the operator $A[\bar\phi,\phi]$ and functional
differentiation.  As discussed in the text following
(\ref{eq:fundamental-relation}), choosing $A=1$ gives the Jarzynski
relation, and choosing $A = e^{(iq+\beta)W}$ leads to the Crooks
relation.

We can obtain a nonequilibrium generalization of the
fluctuation-dissipation relation by differentiating
(\ref{eq:fundamental-relation}) with respect to $U({\bf x},t)$ for
some intermediate time $0<t<t_f$.  From the field theoretic average
(\ref{eq:dp-average}) it follows that the derivative of the left
hand side is
\begin{equation}
  \frac{\delta}{\delta U({\bf x},t)} \langle A e^{-\beta W}\rangle
  = -\biggl\langle \biggl(\beta \frac{\delta W}{\delta U}+
  \frac{\delta S}{\delta U}\biggr)A e^{-\beta W}\biggr\rangle,
\end{equation}
with
\begin{align}
  \frac{\delta W}{\delta U} &=
  \frac{\delta}{\delta U({\bf x},t)} \int_0^{t_f} dt'
  \int d{\bf x}'\, \dot U({\bf x}',t')
  \bar\phi({\bf x}',t')\phi({\bf x}',t') \nonumber\\
  &= -\frac{\partial}{\partial t}\Bigl(\bar\phi({\bf x},t) \phi({\bf x},t)
  \Bigr)
\end{align}
from (\ref{eq:dp-work}), where we have used integration by parts, and
\begin{align}
  \frac{\delta S}{\delta U} &= -\gamma \frac{\delta}{\delta U({\bf x},t)}
  \int_0^{t_f} dt'\int d{\bf x}' \bar\phi \nabla \cdot (\phi\nabla U) \nonumber\\
  &= -\gamma \nabla \cdot \Bigl( \phi({\bf x},t) \nabla \bar\phi({\bf x},t)
  \Bigr)
\end{align}
from (\ref{eq:dp-bulk-action}), where we have used integration by parts
twice.  In contrast, the functional derivative of the right hand side
of (\ref{eq:fundamental-relation}) only acts on the action, yielding
\begin{equation}
  \frac{\delta}{\delta U}\langle \tilde A \rangle_R
  = \gamma \Bigl\langle \nabla\cdot\Bigl(\phi({\bf x},t_f-t) \nabla
  \bar\phi({\bf x},t_f-t)\Bigr)\tilde A
  \Bigr\rangle_R.
\end{equation}
Here we have assumed that $A$ and $A_R$ do not depend on the site
potential $U$.
Combining these gives an identity
\begin{align}
  \frac{\partial}{\partial t}&\langle \bar\phi({\bf x},t)\phi({\bf x},t)
  A e^{-\beta W}\rangle + D\langle \nabla\cdot(\phi({\bf x},t)\nabla
  \bar\phi({\bf x},t) Ae^{-\beta W}\rangle
  \nonumber\\
  &=De^{-\beta\Delta F}\langle \nabla\cdot(\phi({\bf x},t_f-t)\nabla
  \bar\phi({\bf x},t_f-t) A_R\rangle_R,
  \label{eq:noneq-fd-i}
\end{align}
where we have used $\gamma = \beta D$.  The significant of this
identity is that it holds arbitrarily far from equilibrium, i.e., it
is not a linear response relationship.  Additional relations could
be derived by further differentiation with respect to $U$.

For the choice $A = \bar\phi({\bf x}',t')\phi({\bf x}',t')$
Eq.~(\ref{noneq-fd-i}) becomes a nonequilibrium fluctuation
dissipation relation.  Let $\rho = \bar\phi\phi$ and take $U({\bf
  x},t)\to U({\bf x},t) + U_1({\bf x},t)$.  Then
\begin{align}
  \beta&\frac{\partial}{\partial t}\langle \rho({\bf x},t)
  \rho({\bf x}',t') e^{-\beta W} \rangle  \nonumber \\
  &=
  \frac{\delta\langle \rho({\bf x}',t') e^{-\beta W}\rangle}{\delta
    U_1({\bf x},t)}
  - e^{-\beta\Delta F}
  \frac{\delta\langle\rho({\bf x}',t_f-t')\rangle_R}{\delta U_1({\bf x},t_f-t)}
  \label{eq:noneq-fd-ii}
\end{align}
with the functional derivatives evaluated at $U_1=0$, and with the
understanding that $W$ is determined by $U$ and not $U_1$, as defined
in (\ref{eq:dp-work}).  Thus a time derivative of a correlation
function, modified to include the Jarzynski work term, is related to a
response function. The right hand side has effectively a $\theta$
function, with the first term nonzero when $t'>t$ and the second term
nonzero with $t'<t$.

We note that taking functional derivatives of the Jarzynski relation to
obtain a fluctuation dissipation relation within linear response, as
done in \cite{Chetrite2008,Chetrite2009}, differs from this field theoretic
result which, like that obtained in \cite{Mallick2011}, applies arbitrarily
far from equilibrium.

\section{Summary}

\label{sec:summary}

We have generalized the Doi representation to describe interacting
particles coupled to a thermal reservoir, undergoing hops with rates
determined by detailed balance.  Such a system can be driven far from
equilibrium by a rapidly-varying local potential.  We demonstrated that
the Jarzynski and Crooks relations arise straightforwardly in the Doi
representation, with the initial state and the projection state playing
a crucial role.

Then, by mapping the Doi representation to a field theory via the
coherent state representation, we obtained the Doi-Peliti field theory
for these interacting Brownian particles.  We demonstrated that time
reversal in this field theory has a gauge-like form, and that the time
reversal operation results in a fundamental identity
(\ref{eq:fundamental-relation}) from which the Jarzynski and Crooks
relations can be derived, along with a nonequilibrium generalization
of the fluctuation-dissipation relation.

Future work could include generalizing the formalism to include
chemical reactions and allowing for chemical work to be done on the
system, and deriving a general framework for entropy production within
Doi-Peliti field theory.  It could also be fruitful to develop a
perturbative expansion in the interaction strength. 

\begin{acknowledgments}
BPV-L would like to thank Chris Jarzynski and Uwe T\"auber for
helpful discussions.
\end{acknowledgments}


\input{paper.bbl}

\end{document}

%% file: paper.bbl
%